\documentclass[twocolumn,prl,superscriptaddress,floatfix,showpacs]{revtex4-1}
\usepackage[utf8]{inputenc}
\usepackage{amsmath}
\usepackage{amssymb}
\usepackage{graphicx}

\makeatletter
\usepackage{graphics}
\usepackage{epsfig}
\usepackage{color}

\newcommand{\be}{\begin{equation}} \newcommand{\ee}{\end{equation}}
\newcommand{\bea}{\begin{eqnarray}} \newcommand{\eea}{\end{eqnarray}}

\makeatother

\begin{document}

\title{Ubiquity of collective irregular dynamics in balanced networks of spiking neurons}

\author{Ekkehard Ullner} 
\affiliation{Institute for Complex Systems and Mathematical Biology and Department of Physics (SUPA), 
Old Aberdeen, Aberdeen AB24 3UE, UK}
\author{Antonio Politi}
\affiliation{Institute for Complex Systems and Mathematical Biology and Department of Physics (SUPA), 
Old Aberdeen, Aberdeen AB24 3UE, UK}
\affiliation{Max Planck Institut f\"ur Physik komplexer Systeme, N\"othnitzer Str. 38, 01187 Dresden, Germany}
\author{Alessandro Torcini} 
\affiliation{Laboratoire de Physique Th\'eorique et Mod\'elisation, Universit\'e de Cergy-Pontoise, CNRS, UMR 8089,
95302 Cergy-Pontoise cedex, France}
\affiliation{Max Planck Institut f\"ur Physik komplexer Systeme, N\"othnitzer Str. 38, 01187 Dresden, Germany}
\affiliation{Aix Marseille Univ, INSERM, INMED, and INS, Inst Neurosci Syst, 13000 Marseille, France}
\affiliation{CNR - Consiglio Nazionale delle Ricerche - Istituto dei Sistemi Complessi, via Madonna del Piano 10, I-50019 Sesto Fiorentino, Italy}

\date{\today}

\begin{abstract}

We revisit the dynamics of a prototypical model of balanced activity in networks of spiking neurons.
A detailed investigation of the thermodynamic limit for fixed density of connections
(massive coupling) shows that, when inhibition prevails, the asymptotic regime is not asynchronous 
but rather characterized by a self-sustained irregular, macroscopic (collective) dynamics. 
So long as the connectivity is massive, this regime is found in many different setups:
leaky as well as quadratic integrate-and-fire neurons; large and small coupling strength; weak and
strong external currents.
\end{abstract}

\maketitle

{\bf  
Dynamical regimes where excitation and inhibition almost balance each
other are considered very important in computational neuroscience, since 
they are generically accompanied by strong microscopic fluctuations such 
as those experimentally observed in the resting state of the mammalian brain.
While much is known on the balanced regime in the context of binary
neurons and in networks of rate models, much less is known in the more 
realistic case of spiking neural networks.
So far, most of the research activity on spiking neurons was restricted
to diluted networks, with the goal of providing a detailed description 
of the underlying asynchronous regime.
In this paper we show that, contrary to the current expectations, even
in the presence of a 10\% dilution, the collective dynamics exhibited 
is characterized by a sizeable synchronization. The analysis of a
suitable order parameter reveals that the macroscopic dynamics is highly irregular and remains such in the thermodynamic limit (i.e. for infinitely many neurons). 
The underlying form of synchronization is thereby different
from the collectively regular dynamics observed in systems such as the
Kuramoto model.  
}

In spite of the many studies carried out in the last decade, a general theory of the dynamics of 
large ensembles of oscillators is still lacking even for relatively simple setups where the single
units are assumed to be one-dimensional phase oscillators \cite{winfree}.
A whole variety of phases has been indeed discovered which interpolate between the fully synchronous and the
asynchronous regime, including chimera states, self-consistent partial synchrony, not to speak 
of various clustered states \cite{kura,chimera,partial,clustered}.

Even though real systems are composed of a finite number of elements, we know from statistical mechanics
that a meaningful identification of the different regimes can be made only in the thermodynamic limit,
i.e. for an ideally infinite number of elements.
In the case of dynamical systems defined on regular lattices with short range interactions,
taking the limit is straightforward: it is just the matter of considering infinitely extended
lattices. In networks with long-range interactions, the question is less obvious \cite{long}.
Since the interaction grows with the system size, the coupling strength must be inversely proportional
to the number of connections to avoid unphysical divergencies.
Systems like the Kuramoto model belong to this class \cite{kura}.
In setups where the average coupling contribution is negligible, the coupling strength is instead
assumed to scale as the inverse of the square root of the connectivity. 
Spin glasses are the most prominent physical systems where this latter scheme is 
adopted \cite{mezard1987,cris1}.

The characterization of the balanced regime represents another such setup~\cite{cris2} encountered
in computational neuroscience. A theory of balanced states has been developed in ensembles of neurons 
characterized by a coarse-grained variable: their 
firing-rate~\cite{bal1,bal2,bal3,bal4,hansel2015,farzad2017}. However, it is still unclear whether the
resulting scenario is truly representative of what can be observed in more realistic setups.

In fact, an increasing attention has been progressively devoted to simple models of excitatory and 
inhibitory spiking neurons, such as leaky (LIF) or quadratic (QIF) integrate-and-fire neurons with the 
goal of mimicking the cortical 
activity~\cite{gutkin1998, rauch2003,jolivet2004, jolivet2006,shlizerman2012}.  
The most detailed theoretical analysis of spiking neurons
has been proposed by Brunel~\cite{brunel2000}, who derived and solved a (self-consistent) 
Fokker-Planck equation for the probability density of membrane potentials in a network of LIF neurons.
The theory was developed by assuming a finite sparse connectivity, so that the thermodynamic 
limit is implicitly taken by letting the number of neurons diverge.
As the resulting scenario - an asynchronous regime and two kinds of synchronous activity - 
does not fully match the one found in rate models, several numerical studies have been performed 
to investigate the role of ingredients such as the synaptic time scale or the 
network connectivity~\cite{bal2,bal3,rosenbaum2014,bal4,hansel2015}.
The overall result is the evidence of some features which seem to conflict with the hypothesis 
of a widespread existence of a single ``standard" asynchronous dynamics.
For instance Ostojic claims the existence of two different regimes that can be detected
upon increasing the coupling strength \cite{ostojic}. Even though this statement has been challenged 
by Engelken et al. \cite{comment}, who maintain that a single, standard, asynchronous regime does exist, 
the qualitative features of the spiking activity need to be better understood.

In this Article, we revisit the activity of a balanced network of spiking neurons and propose a
different interpretation of the regimes that have been observed in simulations of
finite networks. Our approach is based on a thermodynamic limit,
which better preserves the qualitative features observed in finite systems.
All of our studies show that the network activity is not asynchronous but rather a 
manifestation of a collective 
irregular dynamics (CID), similar to what found in heterogeneous networks of globally coupled 
inhibitory neurons \cite{luccioli2010}.

Collective dynamics can be quantified by implementing the same indicators introduced to measure
the degree of synchronization. With the help of an order parameter specifically
designed to characterize neuronal synchrony in large ensembles of neurons \cite{scholarpedia}, 
we find that CID is an ubiquituous phenomenon, which does not only persist for arbitrary  
coupling strength, but also in the absence of delay and refractoriness. Finally, we find 
that collective dynamics is not restricted to LIF neurons, but extends at least to QIF neurons
as well. All numerical calculations have been performed by implementing either an event-driven 
approach~\cite{zillmer2006,integration} or Euler's algorithm.

We start considering an ensemble of $N$ supra-threshold LIF neurons
composed of $bN$ excitatory and $(1-b)N$ inhibitory cells, as defined in Refs.~\cite{brunel2000,ostojic}. 
The membrane potential $V_i$ of the $i$th neuron evolves according to the equation,
\begin{equation}
\tau \dot V_i = R (I_{0}+I_i) - V_i  \; ,
\label{eq:LIF}
\end{equation}
where $\tau = 20$ms is the membrane time constant, $R I_{0} = 24$mV is an external DC "current",
and $R I_i$ the synaptic current arising from the mutual coupling
\begin{equation}
RI_i = \tau J \sum_n G_{ij(n)} \delta(t-t^{(j)}_n-\tau_d) \; ,
\label{eq:general}
\end{equation}
where $J$ is the coupling strength. 
The synaptic connections among the neurons are random, with a constant in-degree $K$ for each neuron. 
The matrix elements assume the following values: $G_{ij}=1$ ($-g$), if the pre-synaptic neuron 
$j$ is excitatory (inhibitory), otherwise $G_{ij}=0$.
If $V_j$ reaches the threshold $V_{th} = 20$ mV at time $t^{(j)}_n$, two events
are triggered: (i) the membrane potential is reset to $V_r = 10$ mV and $V_j$ is held fixed 
for a refractory period $\tau_r=0.5$ ms; (ii) a spike is emitted and received $\tau_d = 0.55$ ms 
later by the post-synaptic cells connected to neuron $j$. 
All the other parameters are initially set as in Ref.~\cite{ostojic}, namely: $b=0.8$, $K=1000$, $g=5$, 
and $N=10,000$.

\begin{figure}
\begin{centering}
\includegraphics[width=0.23\textwidth,clip=true]{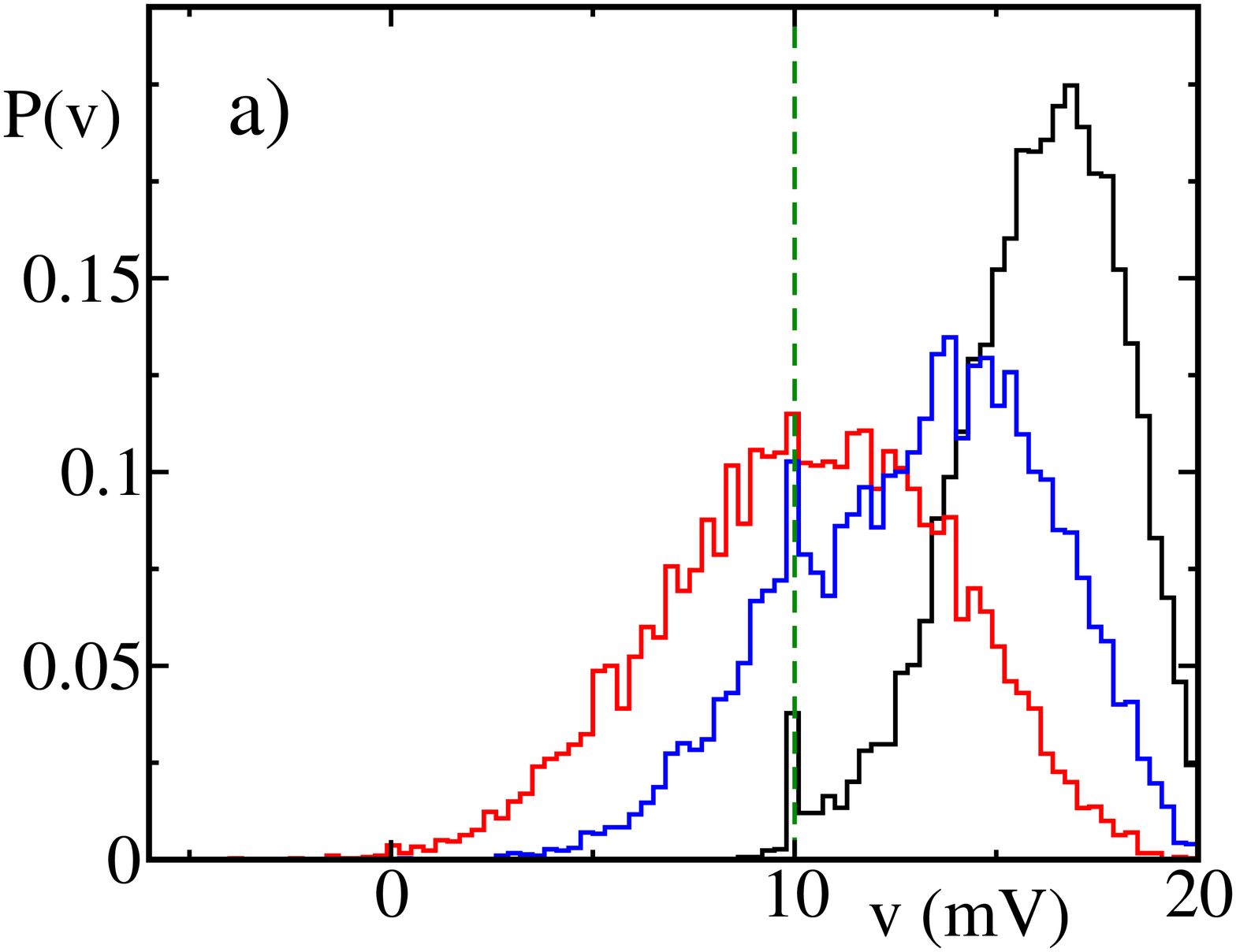}
\includegraphics[width=0.23\textwidth,clip=true]{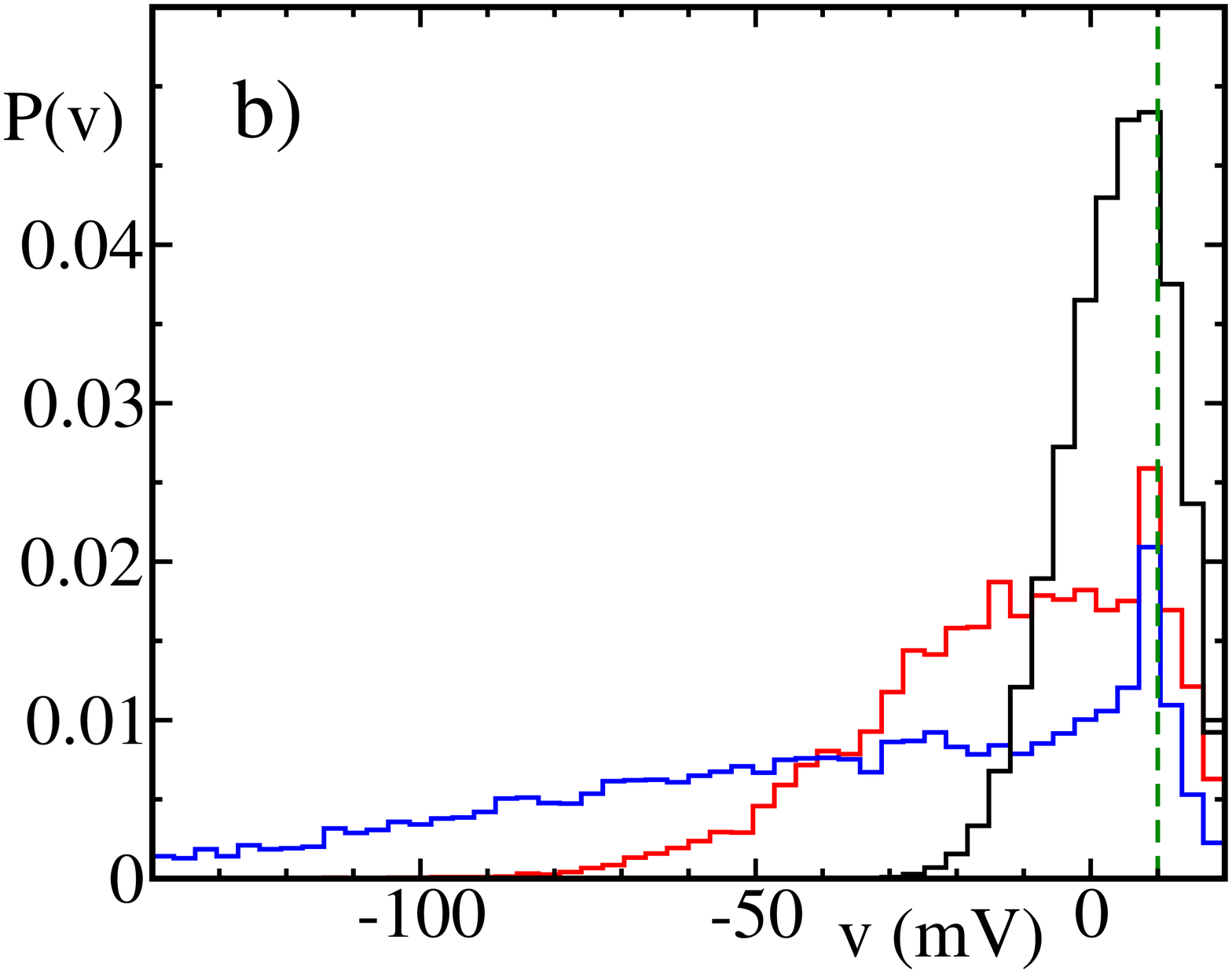}
\end{centering}
\caption{\label{fig:snap}
Three different instantaneous probability distributions of the membrane potentials $P(v)$ for $N=10,000$ and $J=0.1$ mV (a) and $J=0.5$ mV (b). The green dashed line refers to the reset potential $V_r$.}
\end{figure}

We first compute the instantaneous probability density $P(v)$ of membrane potentials $V_i\in[v,v+dv]$ for $J=0.1$ mV and 0.5 mV. 
The asynchronous regime is by definition characterized by a constant firing rate \cite{gerstner} (in the thermodynamic limit). This implies that the flux of neurons along the $v$-axis is independent of both potential and time, i.e. the corresponding probability density $P(v)$ is stationary. From Fig.~\ref{fig:snap}, 
where three different snapshots of $P(v)$ are plotted,
we notice instead strong fluctuations, which appear to grow
with the coupling strength $J$.

Such large fluctuations are inconsistent with the stationarity of the asynchronous regime.
In order to better understand their nature, it is necessary to take the thermodynamic
limit. This can be done in various ways.
In Ref.~\cite{brunel2000}, $N$ is let diverge to $\infty$, keeping all other parameters constant. 
This limit is not able to capture the fluctuations seen in Fig.~\ref{fig:snap}, which indeed
slowly vanish upon increasing $N$. 
In most of the literature on balanced states~\cite{bal1,bal3,bal4,hansel2015}, first the
limit $N \to \infty$, is taken, and then the average in-degree $K$ is let diverge
under the assumption that the coupling strength $J$ is on the order 
of $\mathcal{O}(1/\sqrt{K})$, i.e. one can rewrite explicitly $J = \mathcal{J}/\sqrt{K}$, and $R I_{0} \propto \sqrt{K}$. 
In this Article we propose to let $N$ and $K$ diverge simultaneously, assuming $K=cN$ (this corresponds to assuming a massive connectivity).
A priori, there are two meaningful setups that can be considered: (W) weak external current, 
which corresponds to assume that $R I_{0}$ is independent of $N$ (and thereby $K$);
(S) strong external current, i.e. $I_0 = i_0 \sqrt{N}$.  
In the (W) setup the balance must be ensured a priori by imposing that excitatory and inhibitory fields nearly compensate each other. This is obtained by setting $g \equiv g_0 + g_1/\sqrt{N}$ with $g_0 = b/(1-b)$ so that the average difference between the excitation and inhibition is of the same order as statistical fluctuation. In the (S) setup there is no need to tune $g$ because the external current $RI_0$ maintains the balance. In this Article we show that CID emerges in both setups.

\begin{figure}
\begin{centering}
\includegraphics[width=0.45\textwidth,clip=true]{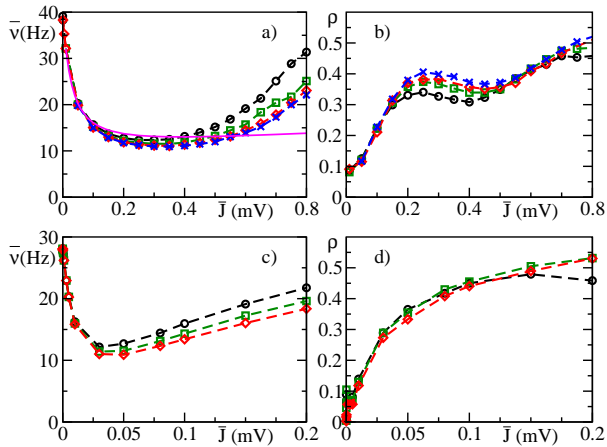}
\end{centering}
\caption{\label{fig:S}
Average firing rate $\overline \nu$ (a,c) and order parameter $\rho$ (b,d) versus the coupling strength $\bar{J}$ for different network sizes $N$: black circles refer to $N=10,000$, green squares to
$40,000$, red diamonds to $160,000$ and blue crosses to $640,000$ (only in (a) and (b)). 
The results refer to the (W) setup for LIF (a,b) and QIF (c,d).
The indicators have been averaged over a time window of $40$ s
and over up to seven realizations of the network.
The solid magenta line (a) is obtained by invoking the diffusion approximation for a noise
driven LIF \cite{ricciardi}.
}
\end{figure}

We first report the results for increasing network sizes for the setup (W), starting from 
$N=10,000$, and including 40,000, 160,000, and 640,000. We set $c=0.1$ and $g_1=100$, as the
resulting model, for $N=10,000$, is equivalent to that in Ref.~\cite{ostojic}.
In Fig.~\ref{fig:S}a we plot the value of the average firing rate $\overline \nu$ versus 
$\bar J =\mathcal{J}/\sqrt{1000}$ \cite{note2}.
In order to damp the (small) sample-to-sample fluctuations, the results are averaged over seven, 
three and two different realizations of the network for 
$N=10,000$, $40,000$ and $160,000$, respectively. We observe a slow but clear 
convergence to an asymptotic curve in the entire range of coupling values.
Finite-size corrections are negligible for $\bar J$ up to 0.1 mV, while for stronger coupling,
the larger the network, the stronger is the tendency of the firing rate to decrease with the system size.
Nevertheless, for $N \gtrsim 160,000$ an asymptotic curve is attained, which exhibits
a growth of $\overline \nu$ with $\bar J$ for sufficiently large coupling (compare with the solid full line,
obtained by invoking the theoretical formula for a noise-driven LIF \cite{ricciardi}).
Another aspect that is maintained in the thermodynamic limit is a bursting 
activity, characterized by a coefficient of variation larger than 1 \cite{izhikevich2007},
for ${\bar J} > 0.3$ mV (data not shown).

A typical order parameter that is used to quantify the strength of collective dynamics
is based on the relative amplitude of the macroscopic fluctuations \cite{scholarpedia}
\begin{equation}
\rho^2 \equiv \frac{\overline{\langle V\rangle^2}-\overline{\langle V\rangle}^2}
    {\langle \overline{V^2}-\overline{V}^2\rangle} \; ,
\end{equation} 
where $\langle \cdot \rangle$ denotes an ensemble average, while the overbar is a time average.
In practice $\rho$ is the rescaled amplitude of the standard deviation of the average $\langle V\rangle$.
When all neurons behave in exactly the same way (perfect synchronization),
the numerator and the denominator are equal to one another and $\rho=1$. If instead, they are
independent, $\rho \approx 1/\sqrt{N}$.
From the results plotted in Fig.~\ref{fig:S}b, we see that the order parameter $\rho$ is finite 
in the whole range of the considered coupling. Furthermore it is substantially
independent of $N$ for ${\bar J} < 0.2$ mV, while for larger ${\bar J}$ it exhibits a slower convergence
to values $\simeq 0.4 - 0.5$. This clearly indicates that the thermodynamic phase is not a standard asynchronous
regime, but is rather characterized by a collective dynamics, also for very small coupling strengths.

\begin{figure}
\begin{centering}
\includegraphics[width=0.45\textwidth,clip=true]{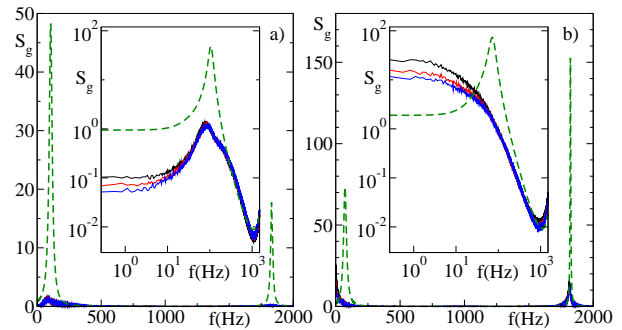}
\end{centering}
\caption{Global spike-train spectra $S_{g}(f)$ versus
the frequency for $\bar J=0.2$ mV(a) and 0.8 mV(b) for LIF in (W) setup.
The different lines refer to different system sizes, namely $N=10,000$ (black), $N=40,000$ (red)
and $N=160,000$ (blue). The dashed green lines show the theoretical results obtained by 
following Ref.~\cite{brunel2000}.
\label{fig:fig4}
}
\end{figure}

The nature of the macroscopic dynamics can be appreciated from the spectrum $\mathcal{S}_g$ 
of the global activity $Y(t)$ (obtained by summing the signals emitted by all the neurons) --
for different system sizes. 
In Fig.~\ref{fig:fig4} we plot the rescaled spectrum $S_g = \mathcal{S}_{g}/N^2$ 
(panels a and b refer to $\bar J=0.2$ mV and $\bar J=0.8$ mV, respectively). 
The data collapse suggests that the dynamics remains irregular in the thermodynamic limit, i.e. that
the fluctuations are not finite-size effects.
In fact, an asynchronous regime would have been characterized by a spectral amplitude $\mathcal{S}_g$ 
of order $\mathcal{O}(N)$ rather than $\mathcal{O}(N^2)$. 
For both coupling strengths the spectral density is mostly concentrated in two frequency ranges: 
(i) around $f \approx 1800$ Hz, which corresponds to the inverse of the delay; 
(ii) at low frequencies in a range that approximately corresponds to the firing rate.
A relative comparison confirms that the collective dynamics is stronger for larger coupling strengths.

Altogether the broad-band structure of the spectrum suggests that the nature of the CID is stochastic-like even though the model is perfectly deterministic. The high-dimensional character of the neural activity is confirmed by a fractal-dimension analysis~\cite{elsewhere,torcini1991}.
To our knowledge we provide the first convincing evidence of collective irregular behaviour in a balanced 
state. The closest regime is reported in a preprint~\cite{hayakawa2017}, which deals with a fully coupled rate model.

The persistence of collective dynamics in the thermodynamic limit in the (W) setup can be
understood in the limit of small connectivity, by revisiting the theory developed by Brunel ~\cite{brunel2000} in the context of highly-diluted networks, 
under the implicit assumption that the thermodynamic limit is taken by letting the number of 
neurons diverge. The central point is the evolution equation 
for the probability $P(v,t)$ 
\begin{equation}
\tau \frac{\partial P}{\partial t} = \frac{\partial}{\partial v} [(v-\mu)P]
+ \frac{\sigma^2}{2} \frac{\partial^2P}{\partial v^2}
+ \sigma_0 \sqrt{c \tau }   \frac{\partial P}{\partial v}\zeta(t)
\label{eq:FP}
\end{equation}
This stochastic Fokker-Planck equation was derived in~\cite{brunel2000} (see Eq.~(32), here
rewritten in our notations); it is valid so long as the current $I$ can be represented as 
the sum of a deterministic contribution $\mu$ and a noise of amplitude $\sigma$.
Self-consistent formulas for $\mu$ and $\sigma$ can be derived
upon assuming an uncorrelated Poisson activity of the various neurons, obtaining 
$\mu(t) = R I_{0} + K J\tau(b-(1-b)g)\nu(t-\tau_d)$, and
$\sigma(t) = J \sqrt{\tau K(b+(1-b)g^2)\nu(t-\tau_d)}$ (see Eqs.~(4,5) in \cite{brunel2000}),
where $\nu(t)$ is the instantaneous firing rate at time $t$ \cite{note3}.

The presence of the common noise in Eq.~(\ref{eq:FP}) is due to the sharing of a fraction
of afferent neurons. In the context of \cite{brunel2000}, the additive noise is a finite-size effect,
since $c=K/N$ vanishes for $N \to \infty$, while in our setup it remains finite.
More precisely, $\zeta(t)$ is a white noise term - $\langle \zeta(t)\zeta(t+T)\rangle= \delta(T)$ - 
while $\sigma_0$ is the value of $\sigma$ corresponding to
the constant firing rate $\nu_0$ obtained within the diffusion approximation \cite{ricciardi}.
By inserting our scaling assumptions for $K$, $J$, and $g$, we find that in the thermodynamic limit
$\mu(t) = R I_{0} -\mathcal{J}\tau \sqrt{c} g_1(1-b)\nu(t-\tau_d)$ and
$\sigma(t) = \mathcal{J}\sqrt{b\tau\nu(t-\tau_d)/(1-b)}$, i.e.
both parameters remain finite when the limit $N\to\infty$ is taken.
As a result, it makes sense to use the solution of Eq.~(\ref{eq:FP}) as a reference for
the results of our numerical simulations.
An analytic expression for the rescaled power spectrum can be found in \cite{brunel2000}
(see the expression reported at the end of page 204). The resulting shape for our parameter
values is reported in Fig.~\ref{fig:fig4} (see the green dashed line in panels a and b).
A qualitative agreement is observed for both $\bar J=0.2$ mV
and $\bar J = 0.8$ mV, starting from the presence of a peak in correspondence of the 
inverse delay. The similarity between the low-frequency peaks is less pronounced for the higher 
coupling strength, showing that the true dynamics is definitely less regular than
theoretically predicted by the noisy Fokker-Planck equation.
The lack of a quantitative agreement is not a surprise, given the perturbative character
of the noisy term and the assumption of a Poisson statistics that is not generally valid.

A closer agreement would be obtained if we could relax some of such approximations.
A promising approach is the self-consistent method developed by Lindner and co-authors~\cite{dummer2014,wieland}, 
which provides a more accurate description of the 
spiking activity of LIF neurons. 
Unfortunately, for our setup above $\bar J = 0.05$ mV the method does not converge~\cite{notelindner},
thus leaving open the question whether it is a technical or conceptual matter.

\begin{figure}
\begin{centering}
\includegraphics[width=0.45\textwidth,clip=true]{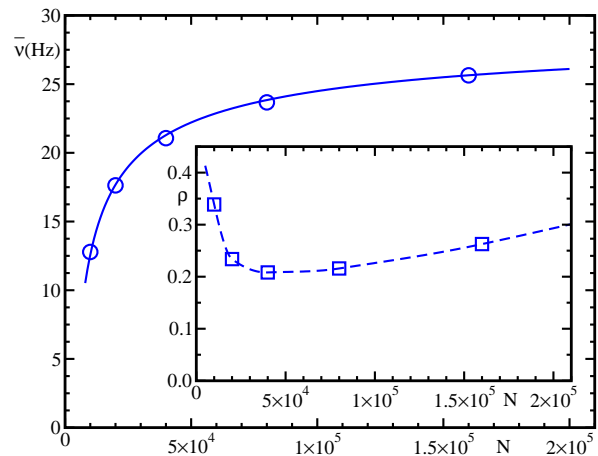}
\end{centering}
\caption{Mean firing rate $\overline \nu$ as a function of the system size $N$ for $\bar J=0.2$ mV
($\mathcal{J}=\bar J \sqrt{1000}$) for the LIF in (S) setup. The solid line, 
$\overline \nu = (30 - 1742.18 / \sqrt{N})$ Hz, shows 
the convergence to the expected asymptotic value $\overline \nu=30$ Hz. 
The inset shows the dependence of the synchronisation measure $\rho$ on $N$ (the dashed line is a guide for the eyes).
\label{fig:strong}
}
\end{figure}

We now discuss the strong-current setup (S), with $R I_0 = 0.24 \sqrt{N}$ mV (i.e.  $R i_0 = 0.24$ mV) and $g=5$. 
In this case, the balance is attained (at leading order in $N$) by imposing the condition
$I_i+I_{0}=0$. Under the assumption of a constant firing rate, this implies 
$\overline \nu = R i_0/(\sqrt{c}\,\tau\mathcal{J}((1-b)g-b)$. The results of simulations for $\overline J=0.2$ mV and
different values of $N$ are reported in Fig.~\ref{fig:strong}, where one can see that the firing rate
converges towards the expected asymptotic value $\overline \nu= 30$ Hz, with a $1/\sqrt{N}$ rate 
(see the solid line). More important is that the order parameter $\rho$ remains finite for increasing $N$ (see
the inset). The presence of strong finite-size corrections prevent us from
determining its asymptotic value; it is however clear that $\rho$ does not vanish, indicating
that a collective dynamics emerges also in the presence of strong external currents.


In order to establish the generality of CID in balanced, massively coupled networks, 
we have analysed another model, the QIF, which represents the canonical model for class I excitability~\cite{gutkin1998,Ermentrout2008,wolf}. Its evolution equation reads as 
\begin{equation}
\tau \dot \theta_i = (1+\cos \theta_i) + (1-\cos \theta_i) (\theta_0 + \alpha RI_i) \; ,
\label{QIF}
\end{equation}
where $\theta_i$ is an adimensional phase-like variable, $\theta_{th}=\pi$ and $\theta_r=-\pi$ are the threshold and reset value, respectively. 
Moreover, $\theta_0= 0.2$, $\alpha = 1\text{mV}^{-1}$, while $RI_i$ is still defined as in Eq.~(\ref{eq:general}),
and all the other parameters are as for the LIF. As shown in Fig.~\ref{fig:S}c,d, where the firing rate $\overline \nu$ and the order
parameter $\rho$ are reported for different coupling strengths, there is again a clear evidence of synchronization. 
The broadband structure of the corresponding spectra of the neural activity (data not shown) indicate that the collective dynamics
is stochastic-like.

Finally, we made several other tests, eliminating refractoriness, setting the delay equal to zero
and adding noise to the external current (as in the original Brunel paper~\cite{brunel2000}). In all these
cases $\rho$ remains finite and exhibits an irregular behavior.

Altogether, we have found that CID emerges in all massively coupled networks we have explored.
This comes as a surprise: in other models of massively coupled neuronal systems, the microscopic
chaotic dynamics which may emerge in finite systems, disappears in the
thermodynamic limit \cite{olmi2010,tattini2012}, while here
it does not only survive but contributes to sustain a macroscopic stochastic-like
evolution. This point definitely needs to be better clarified.

The evidence that CID survives in the vanishing coupling limit could represent the 
starting point for future progress. In fact, for $J=0$ any distribution $P(v)$ is a
marginally stable solution and can in principle be (de)stabilized by an arbitrarily small 
coupling. Such a singular behavior was successfully handled to explain the onset of
partial synchrony~\cite{partial}, by mapping the ensemble of LIF neurons onto the
much simpler Kuramoto-Daido equation~\cite{daido,politi}.
Can one hope to make a similar analysis in the context of the balanced regime?

\begin{acknowledgments}
The authors acknowledge N. Brunel, F. Farkhooi, G. Mato, M. Timme, and M. di Volo for enlightening discussions.
This work has been partially supported by the French government
under the A$\star$MIDEX grant  ANR-11-IDEX-0001-02 (A.T.) and mainly realized at the Max Planck Institute for the Physics of Complex Systems (Dresden, Germany) within the activity of the Advanced Study Group 2016/17 ``From Microscopic to Collective Dynamics in Neural Circuits”.
\end{acknowledgments}


  

\begin{thebibliography}{45}
\bibitem[{not({\natexlab{a}})}]{note2}
\bibinfo{note}{This variable has a one-to-one correspondence with $J$ in
  \cite{ostojic}.}
\bibitem[{els()}]{elsewhere}
\bibinfo{note}{This analysis will be reported in a future publication.}
\bibitem[{not({\natexlab{b}})}]{note3}
\bibinfo{note}{The above equation must be complemented by suitable b.c. to take
  into account refractoriness and the reset.}
\bibitem[{not({\natexlab{c}})}]{notelindner}
\bibinfo{note}{Problems of stability of the method for large inhibition
  (namely, $g=5$) have been previously reported also by the authors in
  \cite{dummer2014}.}
\end{thebibliography}

\end{document}